\documentclass[trackchanges, twocolumn, twocolappendix]{aastex701}

\usepackage{upgreek}

\newcommand{\angstrom}{\textup{\AA} }
\newcommand{\angstromns}{\textup{\AA}}
\newcommand{\mic}{\,$\upmu$m }
\newcommand{\micns}{\,$\upmu$m}
\newcommand{\kms}{\,km\,s$^{-1}$ }
\newcommand{\kmsns}{\,km\,s$^{-1}$}

\newcommand{\nev}{\mbox{[Ne \textsc{v}] }}

\newcommand{\net}{\mbox{[Ne \textsc{iii}] }}
\newcommand{\ot}{\mbox{[O \textsc{iii}] }}

\newcommand{\of}{\mbox{[O \textsc{iv}] }}

\newcommand{\otns}{\mbox{[O \textsc{iii}]}}

\newcommand{\netw}{\mbox{[Ne \textsc{ii}] }}
\newcommand{\netwns}{\mbox{[Ne \textsc{ii}]}}
\newcommand{\artw}{\mbox{[Ar \textsc{ii}] }}

\newcommand{\nath}{\mbox{[Na \textsc{iii}] }}

\newcommand{\art}{\mbox{[Ar \textsc{iii}] }}

\newcommand{\arv}{\mbox{[Ar \textsc{v}] }}

\newcommand{\fetw}{\mbox{[Fe \textsc{ii}] }}
\newcommand{\fetwns}{\mbox{[Fe \textsc{ii}]}}

\newcommand{\st}{\mbox{[S \textsc{iii}] }}

\newcommand{\sfo}{\mbox{[S \textsc{iv}] }}

\newcommand{\sisev}{\mbox{[Si \textsc{vii}] }}

\newcommand{\sinine}{\mbox{[Si \textsc{ix}] }}

\newcommand{\mgeig}{\mbox{[Mg \textsc{viii}] }}

\newcommand{\feeig}{\mbox{[Fe \textsc{viii}] }}
\newcommand{\feeigns}{\mbox{[Fe \textsc{viii}]}}
\newcommand{\mgv}{\mbox{[Mg \textsc{v}] }}

\newcommand{\mgsev}{\mbox{[Mg \textsc{vii}] }}

\newcommand{\fesev}{\mbox{[Fe \textsc{vii}] }}

\newcommand{\nesix}{\mbox{[Ne \textsc{vi}] }}
\newcommand{\nesixns}{\mbox{[Ne \textsc{vi}]}}
\newcommand{\osev}{\mbox{O \textsc{vii} (f) }}
\newcommand{\osevns}{\mbox{O \textsc{vii} (f)}}
\newcommand{\nitsix}{\mbox{N \textsc{vi} (f) }}
\newcommand{\nitsixns}{\mbox{N \textsc{vi} (f)}}
\newcommand{\csix}{\mbox{C \textsc{vi} Ly$\alpha$ }}

\newcommand{\neninens}{\mbox{Ne \textsc{ix} (i)}}
\newcommand{\nitsev}{\mbox{N \textsc{vii} Ly$\alpha$ }}
\newcommand{\nitsevns}{\mbox{N \textsc{vii} Ly$\alpha$}}
\newcommand{\csixB}{\mbox{C \textsc{vi} Ly$\beta$ }}
\newcommand{\csixBns}{\mbox{C \textsc{vi} Ly$\beta$}}
\newcommand{\Cloudy}{\texttt{Cloudy} }

\usepackage{blindtext}
\usepackage{tabularx}

\usepackage{siunitx}
\usepackage{amsmath}
\usepackage{multirow}
\usepackage{enumitem}
\usepackage{xcolor}
\DeclareUnicodeCharacter{2060}{\hspace{0pt}}

\submitjournal{ApJ}

\begin{document}

\title{Unveiling the Multi-Phase Outflow in NGC 4151 with \\ JWST Observations - I. Integrated Properties of the Outflow.}

\author[orcid=0000-0002-5718-2402,sname='Feuillet']{Léa M. Feuillet}
\affiliation{Institute for Astrophysics and Computational Sciences, Department of Physics, The Catholic University of America, Washington, DC 20064, USA}
\email[show]{feuilletl@cua.edu}  

\author[orcid=0000-0003-4073-8977,sname='Kraemer']{Steve Kraemer}
\affiliation{Institute for Astrophysics and Computational Sciences, Department of Physics, The Catholic University of America, Washington, DC 20064, USA}
\email{kraemer@cua.edu}

\author[orcid=0000-0001-8112-3464, sname='Trindade Falcão']{Anna Trindade Falcão}
\affiliation{NASA Goddard Space Flight Center, Greenbelt, MD, 20771, USA}
\email{anna.l.trindadefalcao@nasa.gov}


\author[orcid=0000-0002-3365-8875, sname='Fischer']{Travis C. Fischer}
\affiliation{AURA for ESA, Space Telescope Science Institute, 3700 San Martin Drive, Baltimore, MD 21218, USA}
\email{tfischer@stsci.edu}

\author[orcid=0000-0002-4917-7873, sname='Revalski']{Mitchell Revalski}
\affiliation{Space Telescope Science Institute, 3700 San Martin Drive, Baltimore, MD 21218, USA}
\email{mrevalski@stsci.edu}

\author[orcid=0000-0003-3498-2973, sname='Armus']{Lee Armus}
\affiliation{IPAC, California Institute of Technology, 1200 East California Boulevard,
Pasadena, CA 91125, USA}
\email{lee@ipac.caltech.edu}

\author[orcid=0000-0001-9815-9092, sname='Middei']{Riccardo Middei}
\affiliation{INAF Osservatorio Astronomico di Roma, Via Frascati 33, 00078 Monte Porzio Catone, RM, Italy}
\email{riccardo.middei@inaf.it}

\author[orcid=0000-0002-0982-0561,sname='Reeves']{James N. Reeves}
\affiliation{Institute for Astrophysics and Computational Sciences, Department of Physics, The Catholic University of America, Washington, DC 20064, USA}
\email{reeves@cua.edu}

\author[orcid=0000-0002-2629-4989, sname='Braito']{Valentina Braito}
\affiliation{Institute for Astrophysics and Computational Sciences, Department of Physics, The Catholic University of America, Washington, DC 20064, USA}
\affiliation{INAF, Osservatorio Astronomico di Brera, Via Bianchi 46 I-23807 Merate (LC), Italy}
\affiliation{Dipartimento di Fisica, Università di Trento, Via Sommarive 14, Trento 38123, Italy}
\email{braito@cua.edu}

\begin{abstract}

The biconical outflow of the nearby Seyfert 1.5 AGN NGC 4151 has been studied across the electromagnetic spectrum, yet mapping its multiphase structure with a single facility has remained challenging. Leveraging the broad wavelength coverage of the James Webb Space Telescope (JWST), we present new near- and mid-infrared observations tracing the ionization stratification of these outflows with a unified dataset. We analyze emission lines originating from molecular gas to highly ionized plasma and analyze the ionized gas using a 5-component \Cloudy photoionization model covering three primary phases. Applying the footprint line method to characterize the X-ray emitting gas, we find that it holds the largest share of the total mass, suggesting that it is a primary candidate for driving AGN feedback in this system.

\end{abstract}



\section{Introduction}

Active Galactic Nuclei (AGN) are responsible for some of the most energetic phenomena in the Universe, with the central engine capable of outshining the rest of the host galaxy. The release of vast amounts of energy from the accretion disk around the supermassive black hole (SMBH) results in interaction between the AGN and the interstellar medium of the host. This is commonly referred to as feedback, and often takes the form of winds of ionized gas, which are widespread among the AGN population (e.g. \citealp{Mullaney2013, Woo2016, Torres2020}). Importantly, AGN feedback has been highlighted as one of the necessary phenomena to explain galaxy properties and evolution (e.g. \citealp{Naab2017, Dave2019}) emphasizing the importance of studying these processes in detail. AGN outflows and their role in shaping galaxies have been heavily discussed topics, especially since the 1990s (\citealp{Harrison_twenty_2018}, and references within). This time-frame also corresponds to a noticeable increase in investigations into the outflows of NGC~4151. 


NGC~4151 is a nearby (z = 0.0033, \citealp{deVaucouleurs_redshift_1991}; d = 15.8 Mpc, \citealp{Yuan2020}) Seyfert 1.5 galaxy that has been extensively observed and studied across the full range of the electromagnetic spectrum. The bolometric luminosity of this target has been measured to be L$_{bol}$~=~1.4\,$\times$\,10$^{44}$ erg\,s$^{-1}$, which corresponds to an Eddington ratio of 0.03 \citep{kraemer_mass_2020}, however, the continuum is highly variable (e.g. \citealp{Shapovalova_variability_2010}) which could lead to significant uncertainties in L$_{bol}$. NGC~4151 is also variable in its intrinsic absorption, which has been detected across the X-ray, UV and optical energy ranges (e.g. \citealp{George_variable_1998, kraemer_simultaneous_2005}). Additionally, the nuclear region is notably devoid of star formation (SF) based on the absence of Polycyclic Aromatic Hydrocarbon (PAH) features \citep{Sturm1999}.


Early papers highlighted the archetypal biconical shape of this outflow as seen by Hubble Space Telescope (HST) images \citep{evans_hubble_1993, boksenberg_faint_1995, hutchings_gas_1998}. This was further supported by the available HST spectroscopic data, which demonstrated that the emission lines towards the southwest of the central engine are blueshifted, while those on the northeast side are redshifted \citep{hutchings_gas_1998, kaiser_resolved_2000}. The data also demonstrated the existence of outflowing gas both in emission and absorption at different velocity regimes around $\sim$400\kms and $\sim$1,400\kms \citep{winge_hubble_1999, Nelson2000, crenshaw_space_2000, kraemer_space_2001}.


Early X-ray investigations found extended soft X-ray emission in the same direction as the optically detected bicone \citep{Elvis_extended_xray_1983, Morse_soft_xray_1995}. Spectroscopic investigations into the X-ray emitting regions of NGC~4151 found that the gas is dominated by photoionization \citep{ogle_chandra_2000, armentrout_photoionization_2007}. Early radio investigations of NGC~4151 found radio emission alongside the optical bicone which extends over 950\,pc \citep{Johnston1982}. More recent radio observations at 1.51\,GHz with eMERLIN have found that the emission-line gas further than 360\,pc from the core is consistent with photoionization from the central AGN \citep{Williams2017}.

As demonstrated by the wide range of wavelength regimes through which the outflow in NGC~4151 has been studied, previous research suggests the presence of multi-phase structure in the outflow (spanning a range of density and ionization). Feedback from the lower-ionization components, such as those traced by \ot\,5007\,\angstrom (\ot thereafter), have been accurately measured using spatially resolved kinematic data \citep{crenshaw_feedback_2015, meena_investigating_2023}. However, the absence of X-ray Integral Field Units (IFUs) currently limits our ability to obtain similarly constrained measurements of the X-ray-emitting phase.

A method proposed by \cite{trindade_falcao_tracking_2022} uses "footprint lines", high-ionization emission lines that arise in the same gas producing the X-ray emission, to track the X-ray phase of the outflow. These lines typically have ionization potentials greater than that of O \textsc{vii} (IP\,=\,138\,eV) with some exceptions \citep{feuillet:etal:footprint:2026}. This method has been used to investigate the outflow in NGC~7469, where the authors calculated mass and kinetic energy outflow rates, and found that the most massive and energetic component is the high ionization X-ray gas.

Although both targets are classified as Seyfert 1.5 AGN and have comparable bolometric luminosities (NGC~7469 L$_{bol}$~$\approx$~2.84\,$\times$\,10$^{44}$~erg~s$^{-1}$, \citealp{feuillet:etal:footprint:2026}), the outflow geometry and impact are drastically different. As demonstrated in both \cite{Feuillet2025} and \cite{feuillet:etal:footprint:2026}, the outflow in NGC~7469 does not appear to induce significant globalized feedback into the surrounding ISM, most evident by the presence of a nuclear starburst ring at approximately 500~pc from the AGN. The primary hypothesis is that the bicone is primarily oriented out of the plane of the host galaxy, representing a weak coupling scenario \citep{Harrison2024}. In contrast, the well-defined orientation of the bicone in NGC~4151 in the plane of the host allows us to assume a strong coupling scenario for this target \citep{crenshaw_kinematic_2000, das_mapping_2005}.


This paper is outlined as follows. Section \ref{sec:data} describes the data and data processing methods used in this investigation. Section \ref{sec: modeling} demonstrates the physical basis of our method through photoionization modeling. In Section \ref{sec: outflow}, we explore the multi-phase nature of the outflow through the analysis of the JWST spectrum. In Section \ref{sec:global_modeling}, we use global modeling of the emission lines to estimate mass values for the different components of the outflow. Finally, we summarize and interpret our findings in Section \ref{sec: discussion}.

\section{Observations and Data reduction} \label{sec:data}
\subsection{MIRI and NIRSpec data}\label{sec:IR_data}


An overview of the MIRI and NIRSpec data is given in Table \ref{tab: data}. The MIRI data were taken using the Medium Resolution Spectrometer over a field of view (FOV) up to 6\farcs6\,$\times$\,7\farcs7, and used the FAST readout pattern with 10 groups per integration to avoid saturation and a four-point dither pattern optimized for extended sources. The NIRSpec observations used the high resolution mode and a four-point dither strategy with the small CYCLE dither option to allow the PSF to be subsampled at the 0$\farcs$025 level and offers a FOV of 4\farcs2\,$\times$\,4\farcs8.

\begin{table*}[]
    \centering
    \caption{Summary of JWST and Chandra data reduced and analyzed in this paper.}
    \begin{tabular}{lccccccc}
    \hline
       Telescope & Instrument & $\lambda$/E Range & Channels & Gratings & Exp. Time & Date & Program \\
       \hline\hline
        \multirow{3}{*}{JWST} &\multirow{2}{*}{MIRI} & \multirow{2}{*}{4.9-27.9\mic} & 1, 2, 3, 4 & Short/Medium & 566\,s & 2024-05 & 2773; PI: J. Lyu \\
                     &         &                                 & 1, 2, 3, 4 & Long & 2875\,s & 2024-05 & 2773; PI: J. Lyu \\
        & NIRSpec & 1.66–3.17 \mic & F170LP & G235H & 948\,s & 2022-12 & 1364; PI: M. Bentz \\
        \hline
        \multirow{2}{*}{Chandra} & ACIS-S & 0.3-7\,keV  &  -  & -  & 125\,ks  & 2009-03 & 9217; PI: G. Fabbiano  \\
                                 & ACIS-S & 0.3-7\,keV   &  -  & -  & 75\,ks  & 2009-03 & 9218; PI: G. Fabbiano  \\
        \hline
    \end{tabular}
    \tablecomments{The individual Chandra exposure times are nominal; the combined cleaned exposure used in this analysis is 181.85~ks.}
    \label{tab: data}
\end{table*}

Both the MIRI and NIRSpec data and any associated background data were retrieved from the Mikulski Archive for Space Telescopes (MAST) and processed using the JWST pipeline (v. 1.19.1, \citealp{bushouse_2025}). To conduct a direct comparison between the data from the two instruments, we use the scalexy, ra\_center, and dec\_center parameters set to 0\farcs196, 182.6357191$^{\circ}$, and 39.4058744$^{\circ}$ respectively in the provided pipeline notebooks (\citealp{law_2025_notebooks}). However, despite the forced centering used in both pipelines, the data cubes have a small spatial offset with one another. This is a known issue (e.g. \citealp{Leftley2024, feuillet:etal:footprint:2026}), which we address by aligning the cubes based on their independent continuum peaks.

We additionally process the data using two different alignment options: \texttt{skyalign} and \texttt{ifualign}. As the names suggest, the \texttt{skyalign} alignment outputs cubes which are aligned in the usual coordinate frame, with East pointing to the left and North pointing up, which allows for easy comparison between the IR data and the Chandra data (see Section \ref{sec: Chandra}). On the other hand, \texttt{ifualign} outputs data cubes aligned in the way the data was originally taken by the instrument. As the point spread function (PSF) features are inherently tied to the geometry of the instrument, the PSF should be subtracted from data in this coordinate frame. More information regarding the details of our PSF subtraction process for the JWST data and a discussion on the PSF vs continuum subtraction for the purposes of this investigation can be found in Appendix \ref{app: PSF_sub}.

Our emission line fitting procedure uses \texttt{PySpecKit} to fit the local continuum and gaussian models to the spectroscopic data. A detailed explanation of the method can be found in \cite{feuillet:etal:footprint:2026}. When emission lines originating from the same ion are fit (e.g. \nev 14.32\mic and 24.32\micns), the shorter-wavelength line is first fit, and the Full Width at Half-Maximum (FWHM) and centroid (v$_{cent}$) values are then fixed for the other. The shorter wavelength line is generally less likely to be affected by the fringing issues prevalent in the MIRI data, despite turning on the 1d residual fringe correction in both pipelines.

The uncertainties output from the \texttt{PySpecKit} fitting are not representative of the true uncertainties in fitting our data. This is primarily due to uncertainties representing the formal statistical errors derived from the covariance matrix of the fitting algorithm, without accounting for systematic errors or the limitations of underlying data. To obtain uncertainties in our flux values, we calculate the residuals between the model and the data, which results in higher uncertainties for noisier data, as would be expected. The FWHM uncertainties are primarily due to those in the positioning of the continuum, especially in regions with strong fringing. To estimated the FWHM uncertainties, we calculate the standard deviation in the continuum, fit the line at both the ideal and highest positions of the continuum, and use the difference as the error value. All of the FWHM values reported in the subsequent sections have been corrected for instrumental broadening. Finally, the errors in the v$_{cent}$ values correspond to the instrument's spectral resolution at the wavelength of the emission line. 

\subsection{Chandra/ACIS Data}\label{sec: Chandra}

NGC~4151 has been observed numerous times with the \textit{Chandra X-ray Observatory}. In this work, we use the two deepest observations obtained with the Advanced CCD Imaging Spectrometer spectroscopic array (ACIS-S) in imaging mode without gratings (ObsIDs 9217 and 9218). These observations are included in Chandra Data Collection 656 \dataset[doi:10.25574/cdc.656]{https://doi.org/10.25574/cdc.656}. These observations have nominal exposure times of approximately 125 and 75~ks, respectively, and a combined cleaned exposure of 181.85~ks. This deep ACIS-S dataset has previously been used to study the circumnuclear soft X-ray morphology, emission-line structure, and its relationship with the optical narrow-line region and radio jet \citep{wang_chandra_2011b,wang_chandra_2011a,wang_chandra_2011c}. Here, we reprocess the observations only to generate a soft X-ray image for comparison with the new JWST measurements. We do not perform a new ACIS spectral analysis; the X-ray emission-line measurements discussed in Section~\ref{sec: xray_spectroscopy} and listed in Table~\ref{tab: x-ray lines} are adopted from published high-resolution grating spectroscopy.

The datasets were retrieved from the Chandra Data Archive and reprocessed with \texttt{CIAO} version 4.16 \citep{Fruscione2006} and \texttt{CALDB} version 4.11.3 using \texttt{chandra\_repro}. Standard event filtering was applied, and EDSER subpixel event repositioning was enabled to improve the sampling of the on-axis Chandra point-spread function. Final image processing, including the PSF subtraction described below, was performed with \texttt{CIAO} version 4.17.

\begin{figure}[]
\centering
\includegraphics[width=0.5\textwidth]{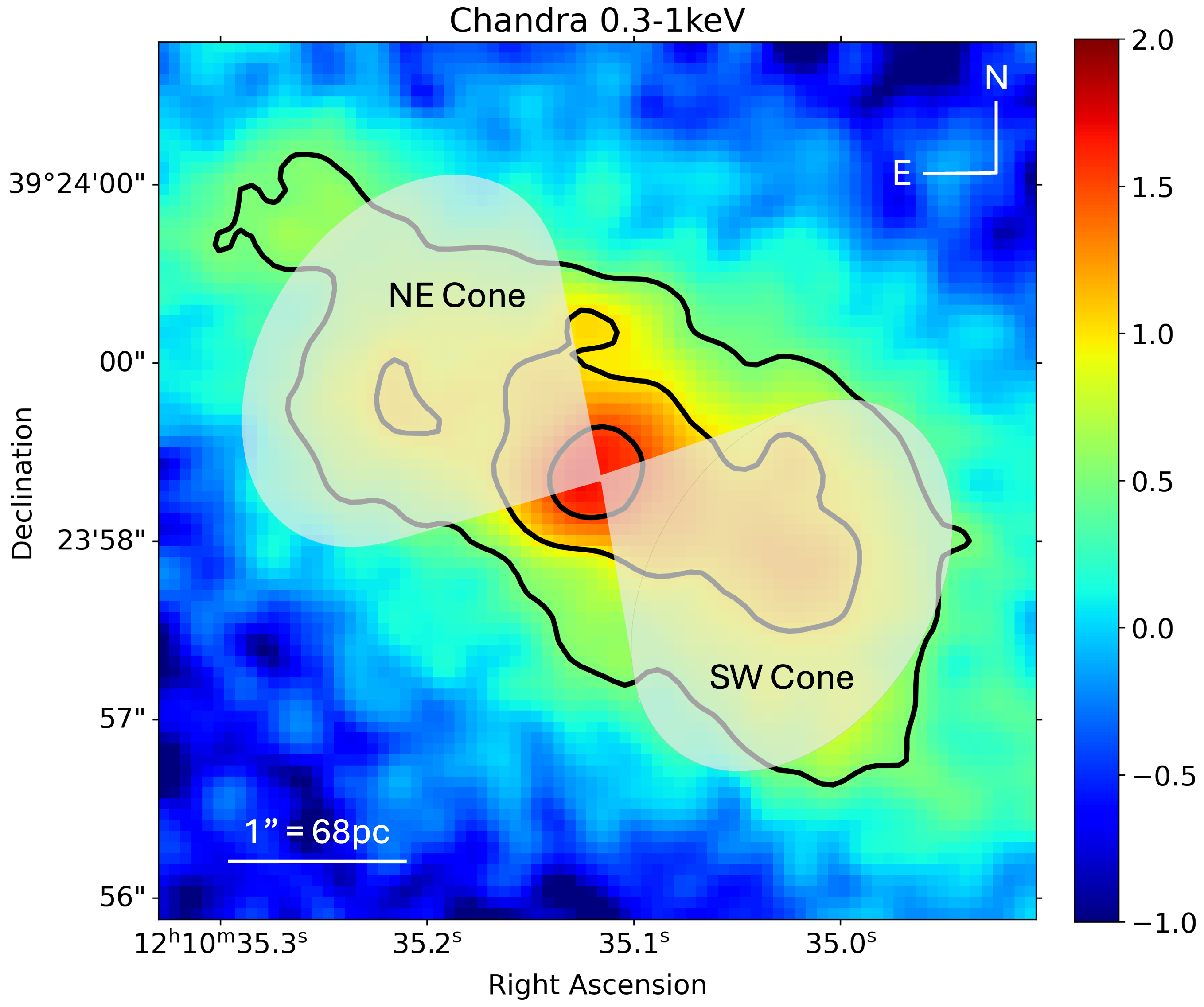}
    \caption{PSF-subtracted Chandra 0.3–1.0~keV image of NGC~4151. The image was smoothed with a Gaussian kernel with $\sigma=1.5$ pixels and a radius of 3 pixels. The contours show the soft X-ray surface-brightness distribution. The modeled ionization bicone from \cite{das_mapping_2005} is overlaid for comparison. The soft X-ray emission shows projected spatial correspondence with the bicone. The southwestern side of the bicone is inclined partly toward the observer, while the northeastern side lies closer to the plane of the sky.
    \label{fig: chandra_soft}}
\end{figure}

ObsID 9217 was adopted as the astrometric reference for the two Chandra observations. Before merging, the world-coordinate system of ObsID 9218 was updated using translations of 0.246 and 0.079 sky pixels along the image x and y directions, respectively, with no rotation or change in scale. The corrected ObsID 9218 event file was then reprojected onto the ObsID 9217 reference frame, and the registered event files were combined. The 0.3-1.0~keV image used in this analysis was sampled on a $0\farcs0615$ pixel grid.

We focus on the 0.3-1.0~keV band, which emphasizes the extended soft X-ray emission. Previous spatially resolved ACIS spectroscopy shows that nuclear photoionization provides the dominant contribution to the large-scale soft X-ray emission in NGC~4151, although the optical and X-ray structures are not identical and localized collisionally ionized emission is present \citep{wang_chandra_2011b,wang_chandra_2011c}. The broadband image used here does not independently separate these emission mechanisms. To separate the extended emission from the bright unresolved nucleus, we generated an observation-specific PSF simulation for each dataset. The PSF simulations were performed using the Chandra Ray Tracer (\texttt{ChaRT}; \citealt{carter_chart_2003})\footnote{\url{https://cxc.cfa.harvard.edu/ciao/PSFs/chart2/index.html}} and projected onto the detector plane using \texttt{MARX}\footnote{\url{https://space.mit.edu/cxc/marx/}}. The simulated event files were centered at the source position and merged in the same energy bands adopted for the imaging analysis.

The merged simulated PSF was aligned with the nuclear position and scaled by a factor of 0.32 to match the observed nuclear peak. The scaled PSF was subtracted pixel by pixel from the 0.3–1.0~keV image using \texttt{dmimgcalc}. The resulting image is used to improve the visibility of the extended circumnuclear emission. Residuals within the central PSF-dominated region remain sensitive to the adopted centering and normalization.

We show the resulting image of the soft X-ray band from 0.3–1.0~keV in Figure \ref{fig: chandra_soft}. The image clearly emphasizes the outflow originating from the nucleus and extending into opposite directions, forming a well-defined biconical structure. We overplot the modeled bicone from \cite{das_mapping_2005}, which was constructed based on HST Space Telescope Imaging Spectrograph (STIS) data, more specifically \ot line emission. We find that the X-ray outflow and the \otns-based model are well-aligned. We also find that the north-east side of the bicone (pointed away from us) is weaker in the X-ray compared to the other side. This is likely due to the intervening gas from the disk of the host and the geometry of the bicone.

\subsection{X-ray Spectroscopic Data}\label{sec: xray_spectroscopy}

Due to the extensive analyses that have been conducted on the X-ray spectroscopy of NGC~4151, both using XMM-Newton's Reflection Grating Spectrograph (RGS) and Chandra's High Energy Transmission Grating (HETG), we do not aim to reanalyze the data. We summarize the values for the lines found in the literature in Table \ref{tab: x-ray lines}. As all of our emission lines of interest were measured using the RGS instrument, we use the values reported in \cite{armentrout_photoionization_2007} for the X-ray line fluxes in following sections. However, for the kinematic information used in Appendix \ref{app: kinematics}, we report the FWHM and v$_{cent}$ values from both RGS and HETG \citep{kraemer_mass_2020}.

\begin{table}[h!]
\centering
\caption{Fluxes, FWHM, and centroid values for the X-ray emission lines.}
\label{tab: x-ray lines}
\resizebox{\columnwidth}{!}{%
\begin{tabular}{lcccccc}
\hline
Line  & IP & Flux & FWHM   & v$_{cent}$ & Ref. \\
 & (eV) & (10$^{-14}$ erg\,cm$^{-2}$\,s$^{-1}$) & (km~s$^{-1}$) & (km~s$^{-1}$) & \\
\hline\hline
 \multirow{2}{*}{\neninens}  & \multirow{2}{*}{239} & 8.14$\pm$0.85 & -  & -398$^{+99}_{-133}$ & (1) \\
                            &                     & 10.5$\pm$1.0 &  $<$780  & - & (2)       \\
 \hline
 \multirow{2}{*}{\osevns}  & \multirow{2}{*}{138} & 40.4$\pm$1.6 & -  & -321$^{+0}_{-54}$ & (1) \\
                         &                      & 34.0$\pm$7.6 & 500-2500 & -236$^{+199}_{-412}$ & (2)\\
 \hline
 \nitsevns & 552 & 6.82$^{+0.64}_{-0.62}$ & 325$^{+141}_{-150}$  & -240$\pm$60  & \multirow{3}{*}{(1)} \\
 \csixBns  & 392 & 1.75$^{+0.35}_{-0.18}$ & -  & -275$^{+69}_{-137}$ & \\
\nitsixns & 98   & 7.18$\pm$0.56 & 182$^{+75}_{-81}$    & -214$^{+72}_{-71}$  & \\
\hline
\end{tabular}}
\tablecomments{(1) \cite{armentrout_photoionization_2007} (RGS), (2) \cite{kraemer_mass_2020} (HETG).}
\end{table}

\section{Photoionization Modeling} \label{sec: modeling}

\begin{figure*}[]
\centering
\includegraphics[width=\textwidth]{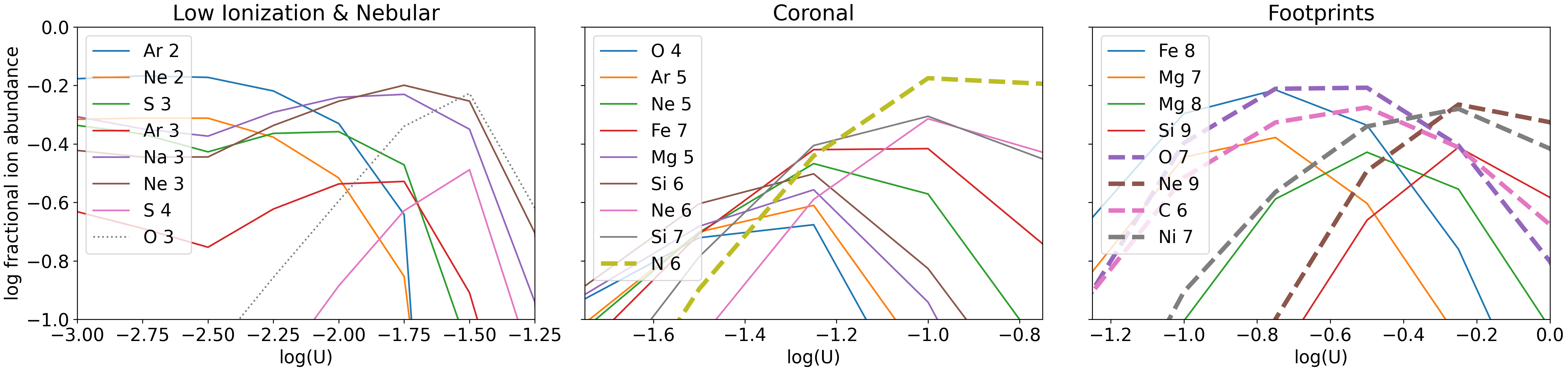}
    \caption{Ionic abundances for the ions corresponding to the detected emission lines in the 1D spectrum of NGC~4151 (Figure~\ref{fig: full_spec}), using a column density of log(N$_{H}$/cm$^{-2}$)\,=\,21.5. We separate the emission lines into three categories: low-ionization/nebular, coronal, and footprint, based on the log(U) value at which the fractional ion abundance peaks. The solid, dashed, and dotted lines represent ions that correspond to IR, X-ray, and optical emission lines of interest respectively. 
    \label{fig: abundances}}
\end{figure*}

As per previous studies of NGC~4151 \citep{armentrout_photoionization_2007, kraemer_mass_2020} and our application of the footprint method \citep{trindade_falcao_tracking_2022, feuillet:etal:footprint:2026}, we use the photoionization code \texttt{Cloudy} (v23.01, \citealp{Cloudy2023}) to obtain the fractional ion abundance as a function of ionization parameter (U) for our emission lines of interest. We define U as
\begin{equation}\label{eq: U}
    U=\frac{Q}{4 \pi  r^{2} c n_{H}}
\end{equation}
where Q is the number of ionizing photons per second emitted by the central object (Q~=~3$\times$10$^{53}$ photons\,s$^{-1}$, \citealp{kraemer_mass_2020}), r is the distance between the central object and the cloud, c is the speed of light, and n$_{H}$ is the hydrogen number density \citep{Osterbrock2006}. We use a power-law spectral energy distribution (L$_\nu$ $\propto$ $\nu^{-\alpha}$) specifically tailored to NGC~4151, used in \cite{kraemer_mass_2020} and included here for easy reference:
\begin{equation*}
        \alpha = 1.0 \textrm{ for } 1 \times 10^{-4} \textrm{ eV}  \leq \textrm{h}\nu \leq 13.6 \textrm{ eV;}
\end{equation*}
\begin{equation*}
        \alpha = 1.3 \textrm{ for } 13.6 \textrm{ eV}  \leq \textrm{h}\nu \leq 500 \textrm{ eV;}
\end{equation*}
\begin{equation*}
        \alpha = 0.5 \textrm{ for } 500 \textrm{ eV} \leq \textrm{h}\nu \leq 30 \textrm{ keV.}
\end{equation*}

Similarly to previous studies \citep{trindade_falcao_tracking_2022, feuillet:etal:footprint:2026}, we use a \Cloudy model\footnote{The ionic fraction plots are not sensitive to density \citep{trindade_falcao_tracking_2022, feuillet:etal:footprint:2026}.} with a column density of log(N$_{H}$/cm$^{-2}$)\,=\,21.5, a hydrogen density of log(n$_{H}$/cm$^{-3}$)\,=\,4 and 1.4$\times$ solar abundances \citep{Asplund2021}. We then vary U between log(U)\,=\,-3 and 0.5 in steps of 0.25 to generate the distributions shown in Figure \ref{fig: abundances}.

The line categories used in the next section are based on where the ion corresponding to each emission line occurs. Low-ionization/nebular lines peaks at or below where the optical \ot line peaks (log(U) $\leq$ -1.5). The coronal lines peak within -1.5 $<$ log(U) $\leq$ -1.0. Finally, the footprint lines peak where the X-ray emission lines, which probe some of the most highly ionized gas in AGN, also peak. Note however, that we are comparing the ionic abundances, which are not affected by the mechanism through which the corresponding emission lines are produced. However, emission line strength would be affected by the X-ray emission lines and IR lines being produced by two different processes (recombination \&  collision respectively). Additional details about the categories defined above will be given in the following sections.

\section{The Multi-Phase Outflow} \label{sec: outflow}

\begin{figure*}[]
\centering
\includegraphics[width=\textwidth]{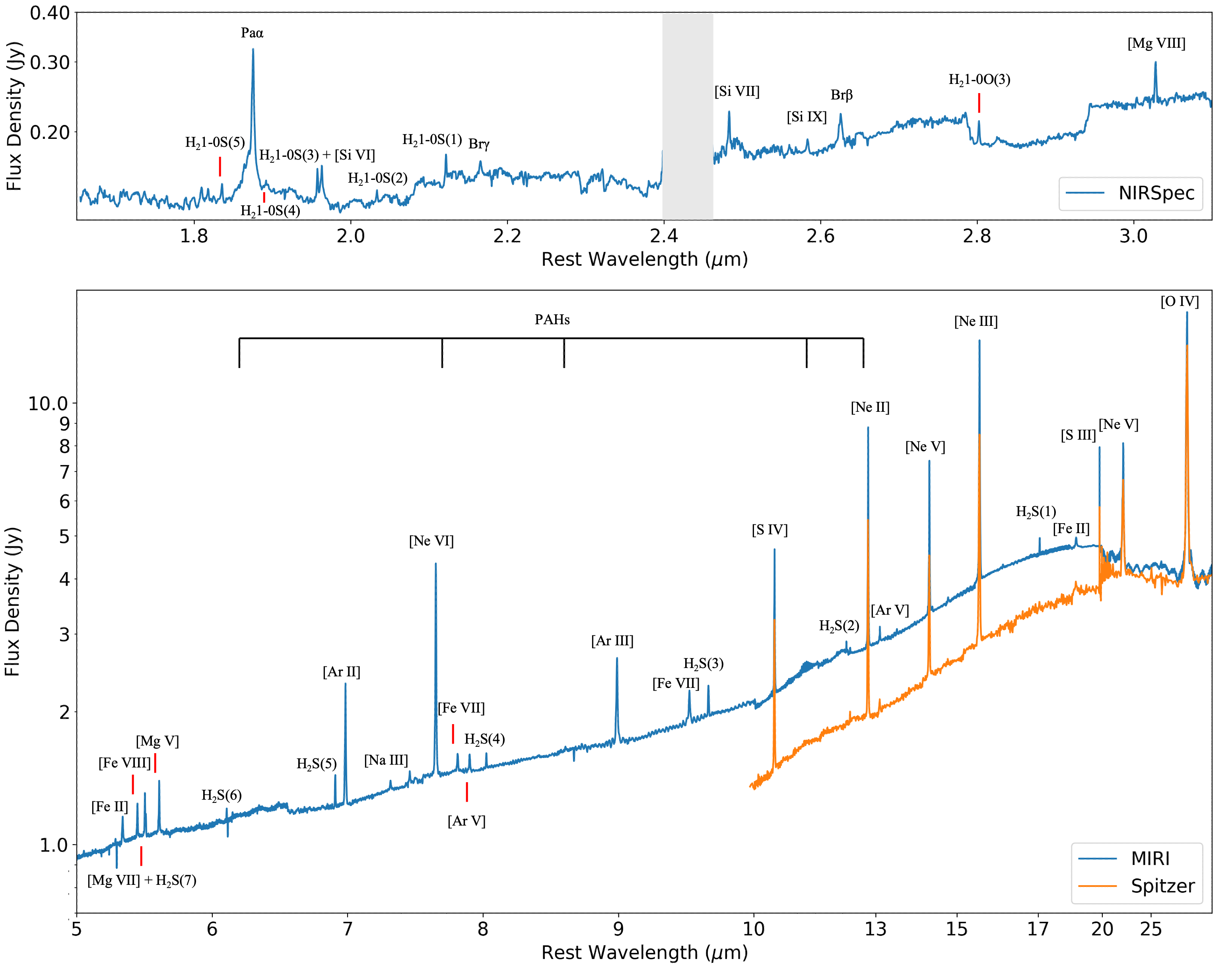}
    \caption{Available total integrated NIRSpec, MIRI and Spitzer spectra of NGC~4151 over each respective FOV. The NIRSpec F170LP/G235H grating is shown in the top panel, while we compare the MIRI and Spitzer spectra in the bottom panel. The more prominent emission lines are labeled. We adjust the scaling on the x-axis to best demonstrate the full range of emission lines in the spectrum. The grey band gap is due to the physical gap between the detectors.
    \label{fig: full_spec}}
\end{figure*}

In Figure \ref{fig: full_spec} we show the full available 1D spectrum for NGC 4151 from the pipeline, alongside the Spitzer spectrum. The optimal extraction of the high resolution Spitzer spectrum (PI: J. Houck) was downloaded from the Combined Atlas of Sources with Spitzer IRS Spectra (CASSIS). The difference between the Spitzer and MIRI spectra could be due to the difference in sensitivity between the instruments, however, our \net value is the same as that measured from the Spitzer spectrum (2.48\,$\pm$\,0.03 and $2.50\,\pm\,0.11\times\,10^{-12}$~erg\,cm$^{-2}$\,s$^{-1}$; from this paper and  \citealp{Deo2007} respectively). This suggests the difference stems from the intrinsic variability of NGC~4151 (e.g., \citealp{Shapovalova_variability_2010}), as the observations span ~20 years. The two infrared continua differ by a factor of ~2, with a stronger variation at shorter wavelengths dominated by hotter dust. Unchanged emission lines indicate the continuum variation could be driven by reprocessed UV radiation.


The subsequent sections examine the physical properties of the distinct gas phases identified in the nuclear region of NGC~4151.

\subsection{Warm Molecular Gas}

The spectrum notably features the H$_2$ rotational line series from H$_2$S(1) to H$_2$S(7), which are believed to originate from the heating of the gas by X-ray photons escaping the central dusty torus \citep{may_nuclear_2020}. Other rovibrational molecular hydrogen lines are also present in the NIRSpec spectrum, such as H$_2$1-0O(3) and H$_2$1-0S(1) to (5). All of the H$_2$ lines are fit using a single gaussian component with FWHM values ranging from 104 to 341\,km\,s$^{-1}$ and an average value of 217\,km\,s$^{-1}$. The centroids are all consistent with systemic (v$_{cent}$~=~0\,km\,s$^{-1}$) when considering uncertainties (Section \ref{sec:IR_data}).

Notably, the molecular hydrogen lines are not detected with a high S/N in the Spitzer spectrum, and are beyond the wavelength range of other IR IFUs used to investigate the nuclear region of NGC~4151. However, the near-IR rovibrational line fluxes were calculated in \cite{Storchi2009} with the Gemini Near-infrared Integral Field Spectrograph. The authors found that the ortho-to-para ratio is indeed approximately 3, and a temperature of T = 2155\,K for the H$_2$. For the total emitting area, the H$_2$ 2.12 flux calculated was 2.5\,$\times$\,10$^{-14}$\,erg\,cm$^{-2}$\,s$^{-1}$, consistent with our own measured value of (3.03\,$\pm$\,0.73)\,$\times$\,10$^{-14}$\,erg\,cm$^{-2}$\,s$^{-1}$. We can use the pure rotational lines of H$_2$ to estimate the total molecular gas mass in the central region of this target using the method described in \cite{Togi2016}. 

A Python code, \texttt{h2powerlaw}\footnote{\url{https://h2powerlaw.readthedocs.io/en/latest/background.html}}, was developed in \cite{Jones2025} and released to the community which implements the \cite{Togi2016} H$_2$ power-law fitting method. We use this software in order to perform our fit, using the total H$_2$ flux values and their corresponding uncertainties as inputs (see Table \ref{tab:molecular}). The code assumes local thermodynamic equilibrium and a fixed ortho-to-para abundance ratio at any temperature.

We obtained a best fit value of 3.81 for the power-law index and a lower temperature of 140\,K, which corresponds to a total H$_2$ mass of $\sim$4.7\,$\times$\,10$^5$\,M$_{\odot}$ between 200 and 2000\,K, and $\sim$1.8\,$\times$\,10$^7$\,M$_{\odot}$ between 50 and 2000\,K. The H$_2$ gas mass was derived in \cite{Storchi2009} and found to be of 240\,M$_{\odot}$. However, the mass was derived for a single high temperature, and therefore does not sample the same range of gas temperatures that we do in our calculation, hence the discrepancy. On the other hand, our values are in good agreement with the maximum warm H$_2$ mass value obtained by \cite{Sturm1999} (3\,$\times$\,10$^7$\,M$_{\odot}$) using Infrared Space Observatory data.

The power-law method was also used to measure the H$_2$ mass in NGC~7469 in the nuclear region only, with a resulting index of 4.55, $\sim$1.3\,$\times$\,10$^5$\,M$_{\odot}$ of warm gas at T $>$ 200\,K and an extrapolated mass down to 50\,K of $\sim$1.2\,$\times$\,10$^7$\,M$_{\odot}$ \citep{armus_goals-jwst_2023}. Given their similar bolometric luminosities and the absence of SF in their inner regions, these values align well with those derived for the nuclear spectrum of NGC 4151. By comparison, \cite{lai_goals-jwst_2022} measured a substantially higher warm H$_2$ mass of (1.2\,$\pm$\,0.3)\,$\times$\,10$^7$\,M$_{\odot}$ in the distinct SF ring of NGC 7469 at T $>$ 200\,K, which samples a much larger region compared to the other measurements.

\begin{figure}[]
\centering
\includegraphics[width=0.5\textwidth]{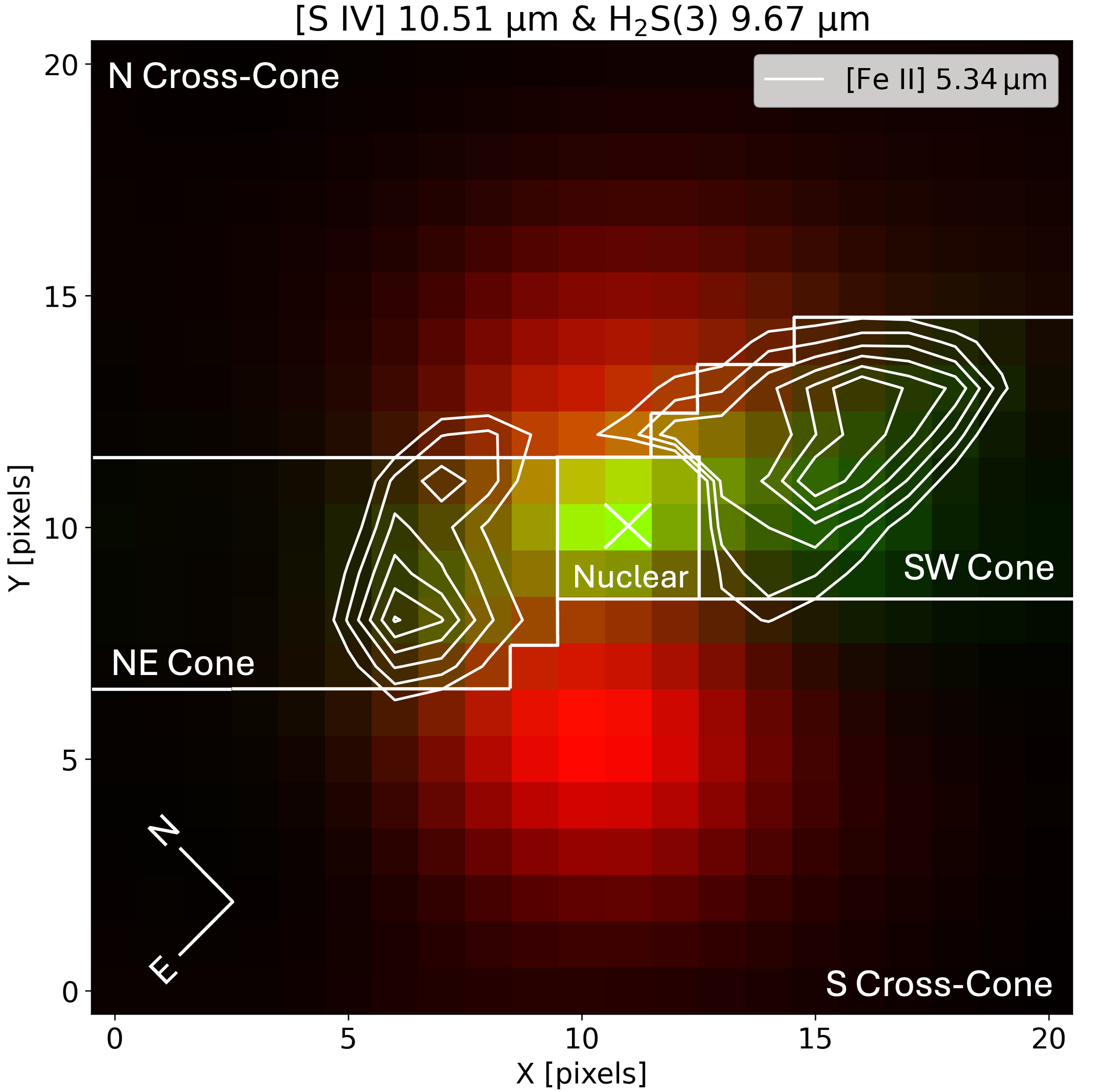}
    \caption{This composite image shows the distribution of the \sfo (green) and H$_2$S(3) (red) gas in NGC~4151. The regions where the \sfo is dominant are defined as the North-East and South-West cones, while the regions dominated by the H$_2$S(3) molecular gas are named the North and South cross-cones. We also plot the contours for the \fetw 5.34\mic emission-line, which is spatially coincident with the ionized outflow bicone.
    \label{fig: regions}}
\end{figure}

\begin{figure}[]
\centering
\includegraphics[width=0.45\textwidth]{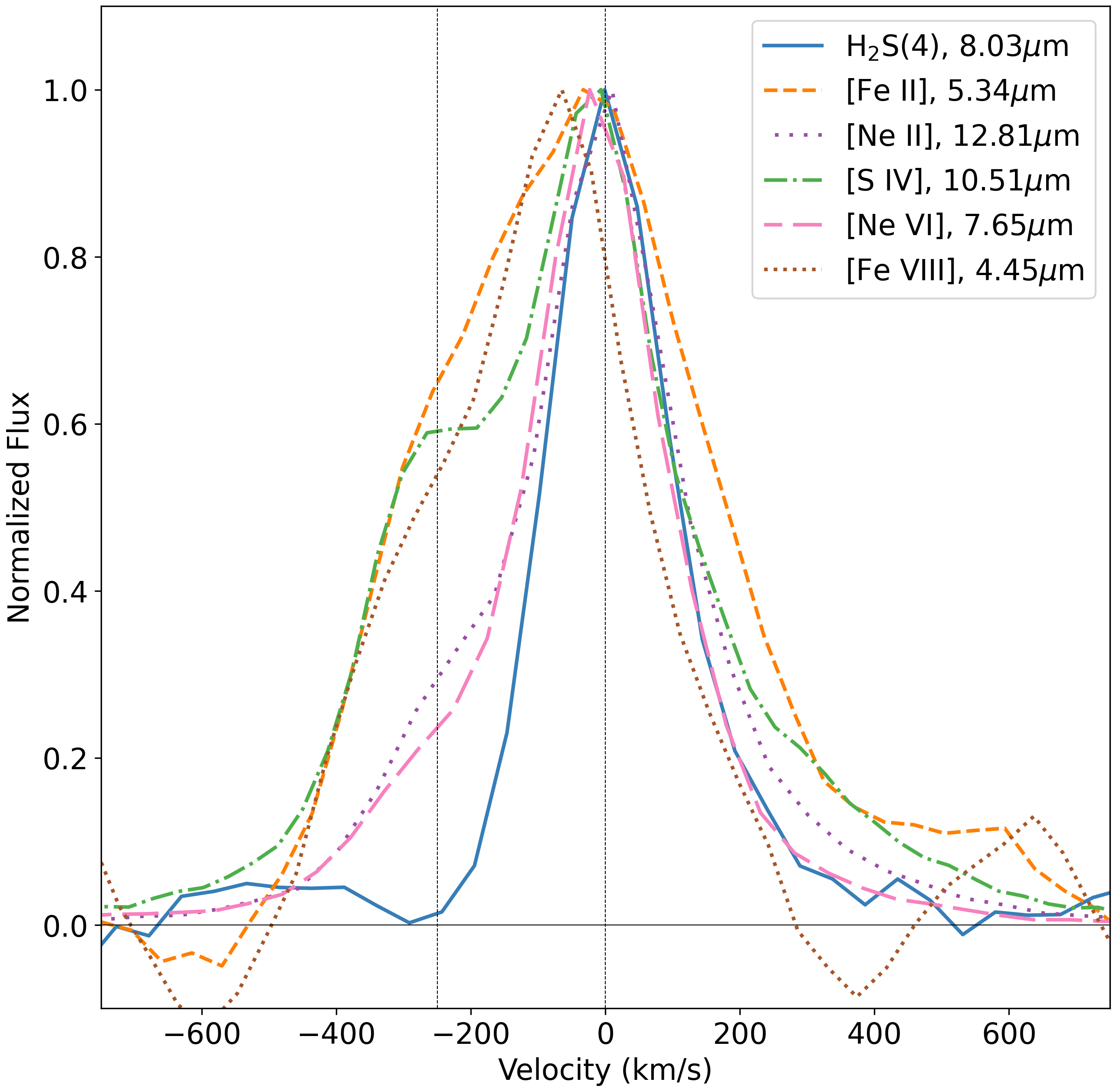}
    \caption{Profile comparison between the strongest emission lines in each category. These profiles were extracted from all regions but the nuclear region as defined in Figure \ref{fig: regions}. The vertical lines indicate velocities of 0\,km\,s$^{-1}$ and -250\,km\,s$^{-1}$.
    \label{fig:profile_comparison}}
\end{figure}

The spatial distribution of the warm molecular gas and \sfo 10.51\mic is highlighted in Figure \ref{fig: regions}. We define five regions: the nuclear region, the NE and SW cones, and the N and S cross-cones. The \sfo is dominant in the cone regions, while the H$_2$S(3) line emission is strongest outside of the AGN bicone, consistent with other AGN (e.g. NGC~5506; \citealp{Delaney2025}, or NGC~1068; \citealp{Riffel2014}). We also plot the line profile for H$_2$S(4), excluding the previously defined nuclear region, in Figure \ref{fig:profile_comparison} for easy comparison with the other line types in subsequent sections. 

The H$_2$ emission is concentrated primarily in the cross-cone region and shows comparatively narrow, near-systemic profiles, with no compelling evidence that the observed warm molecular gas participates directly in the AGN-driven outflow. This is consistent with molecular gas rotating into the illuminated bicone, where it may be dissociated and ionized before being accelerated. However, we note that localized or faint high-velocity molecular components on smaller spatial scales could be masked due to the global extraction.

\begin{center}
\begin{table}[htbp]
\centering
\caption{Summary of the molecular hydrogen lines: best-fit fluxes and FWHM.}
\label{tab:molecular}
\resizebox{\columnwidth}{!}{%
\begin{tabular}{lccc}
\hline
Line & $\lambda_{rest}$ & Flux & FWHM \\
 & (\mic) & (10$^{-14}$ erg\,cm$^{-2}$\,s$^{-1}$) & (km\,s$^{-1}$) \\
\hline\hline
\multicolumn{4}{l}{MIRI H$_2$ Lines} \\
\hline 
H$_2$S(8) & 5.05  &  1.22(0.28) & 232(18) \\
H$_2$S(7) & 5.51  & 4.65(0.74) & 198(7) \\
H$_2$S(6) & 6.11  & 2.55(2.74) & 186(21) \\
H$_2$S(5) & 6.91  & 6.89(0.95) & 203(11) \\
H$_2$S(4) & 8.03  & 3.24(0.37) & 207(15) \\
H$_2$S(3) & 9.67  & 7.54(0.59) & 203(10) \\
H$_2$S(2) & 12.28 & 2.83(0.62) & 206(23) \\
H$_2$S(1) & 17.04 & 4.50(1.10) & 171(12) \\
\hline
\multicolumn{4}{l}{NIRSpec H$_2$ Lines} \\
\hline
H$_2$1-0S(5) & 1.84 & 2.61(0.45) & 341(28) \\
H$_2$1-0S(4) & 1.89 & 0.34(0.09) & 103(36) \\
H$_2$1-0S(3) & 1.96 & 4.92(1.08) & 337(42) \\
H$_2$1-0S(2) & 2.03 & 1.12(0.19) & 239(34) \\
H$_2$1-0S(1) & 2.12 & 3.03(0.73) & 228(13) \\
H$_2$1-0O(3) & 2.80 & 2.00(0.22) & 180(6) \\
\hline
\end{tabular}}
\tablecomments{Uncertainties in the values are given in parentheses.}
\end{table}
\end{center}

\subsection{Hydrogen Recombination Lines}

The Hydrogen recombination lines offer insight into the Broad Line Region (BLR) of NGC~4151. The strongest line, Pa$\alpha$, has not been observed by any instruments in the past for this target. We fit the Pa$\alpha$ line with three components, representing the contribution from the Narrow Line Region (NLR), Intermediate Line Region (ILR; \citealp{crenshaw_mass_2007}), and the BLR. We find that the components have FWHM values of 100, 550 and 3250\,km\,s$^{-1}$, with centroids near systemic for the first two, and a blueshift of -500\,km\,s$^{-1}$ for the broad component. Another study of NGC~4151 in the UV fit the C \textsc{iv} line with three components of FWHM 250, 1,170, and 10,000\,km\,s$^{-1}$ \citep{crenshaw_mass_2007}. These values are likely different from the Pa$\alpha$ fits as C \textsc{iv} is a wind line, and thus is likely to have a contribution from the AGN wind (or mass outflow) compared to the Pa$\alpha$ line originating from the virialized gas.


\subsection{Low-Ionization/Nebular Gas}

Low-ionization emission lines are commonly associated with SF (e.g. \netw 12.81\micns, \st 18.71\micns, \citealp{Feuillet2025}) but can also be produced in AGN-ionized gas. This is noted, particularly for \netwns, in \cite{Melendez2008-contribution}. In NGC~4151, the contribution to these lines from SF is likely to be negligible. Indeed, based on the global fluxes obtained in this paper for the relevant emission lines, this region of NGC~4151 would fall firmly in the AGN region of mid-IR diagnostic diagrams\footnote{We will address the possibility of small SF regions contained within the considered FOV in the second paper of this series by plotting each spaxel on IR diagnostic diagrams.} \citep{Feuillet2025}. Additionally, there is a distinct lack of Polycyclic Aromatic Hydrocarbon (PAH) features in the spectrum (Figure \ref{fig: full_spec}), which has been attributed to the lack of SF \citep{Sturm1999}. Spatially resolved JWST NIRSpec observations further support this, revealing that the emission features and continuum emission across individual spaxels are directly linked to jet interactions, ionization bicone, and H$_2$ emission rather than SF \citep{Chambon2024}.

Similarly to H$_2$ emission lines, \fetw can be produced in partially ionized regions due to X-ray photons. These lines can also be produced in shocks due to the presence of a radio jet or shocks from supernova remnants \citep{Forbes1993, Simpson1996}. As previously established, NGC~4151 has been shown to host a radio jet (e.g. \citealp{Williams2017}) and the peaks in \fetw emission appear to be near the lobes of the radio jet, which suggests that the emission originates from the jet/ISM interaction \citep{may_nuclear_2020}. As shown in Figure \ref{fig: regions}, the \fetw\,5.34\mic emission is not spatially coincident with the warm molecular, H$_2$, gas. Considering the high mass outflow rates ($\dot{M}$ $\sim$ 1-3 M$_\odot$yr$^{-1}$) calculated for this target in the past (e.g. \citealp{Storchi2010}, \citealp{crenshaw_feedback_2015}), this could contradict recent results by \cite{Delaney2026}, in which the authors found that the \fetw emission was more strongly associated with H$_2$ temperatures for targets with higher $\dot{M}$.

Similar to the H$_2$ lines, the \fetw lines are well-modeled with a single gaussian but with higher FWHM values, as can be seen in the Figure \ref{fig:profile_comparison} profiles. This is indicative of disturbances in the gas, which could be due to shocks\footnote{A map of the \fetwns/\artw 6.99\micns, as suggested by \cite{Delaney2026}, would support the possible presence of shocks. This will be addressed in the second paper of this series.} (from the AGN outflow or the radio jet). 

The other lines, listed in Table \ref{tab:low_lines}, are best fit by two Gaussian components: a narrow component and a broad component, both located near systemic. The presence of both narrow and broad emission is consistent with outflowing gas, whereas the lack of a significant net velocity shift suggests that both redshifted and blueshifted gas contribute to the observed emission, as expected from opposite sides of the outflow.

We differentiate the low ionization lines as defined above from nebular lines, which, as defined in \cite{feuillet:etal:footprint:2026}, are emission lines originating from the same \otns-emitting gas commonly studied in the optical. Based on Figure \ref{fig: abundances}, these lines are \sfo\,10.51\micns, \net\,15.56\micns, and \nath\,7.32\micns. However, due to the lack of SF in this target, the nebular lines and other low ionization lines do not exhibit strong differences in terms of their global kinematics (see Figure \ref{fig:profile_comparison}). However, this may not be the case when comparing smaller spatial extractions.

\begin{table}[]
\centering
\caption{Summary of the H recombination lines, low ionization and nebular lines: best-fit fluxes and FWHM.}
\label{tab:low_lines}
\resizebox{\columnwidth}{!}{%
\begin{tabular}{lcccc}
\hline
Line & $\lambda_{rest}$ & IP  & Flux & FWHM \\
 & (\mic) & (eV) & (10$^{-14}$ erg\,cm$^{-2}$\,s$^{-1}$) & (km\,s$^{-1}$) \\
\hline\hline
\multicolumn{5}{l}{H Recombination Lines} \\
\hline
Pa$\alpha$ & 1.88 &  -  & 3.95(0.54)  & 100 \\
 &      &    & 35.10(4.81)  & 550 \\
 &      &    & 61.31(8.39)& 3250 \\
Br$\gamma$& 2.17 &  -  & 1.21(0.16)  & 270  \\
 &      &    & 1.99(0.27) & 800  \\
Br$\beta$  & 2.63 &  -  & 0.73(0.13) & 170  \\
 &      &    & 6.15(1.11) & 580  \\
\hline
\multicolumn{5}{l}{Low Ionization Lines} \\
\hline 
\fetw  & 5.34  & 7.9    & 12.6(1.0)      & 470(31) \\
\fetw  & 17.94 & 7.9    & 7.0(0.3)       & 460(27) \\
\artw  & 6.99  & 15.8   & 11.6(0.3)      & 140(5) \\
        &       &       & 39.8(0.4)      & 460(10) \\
\netw  & 12.81 & 21.6   & 50.4(1.8)      & 184(3) \\
        &       &       & 107.0(3.7)     & 490(18) \\
\st     & 18.71 & 23.3  & 11.7(0.5)      & 186($<$1) \\
        &       &       & 72.3(0.7)      & 537(7) \\
\art    & 8.99  & 27.6  & 10.2(0.1)      & 180(9) \\
        &       &       & 30.2(0.7)      & 469(10) \\
\hline
\multicolumn{5}{l}{Nebular Lines} \\
\hline
\sfo    & 10.51 & 34.8  & 11.89(0.96)     & 122(1) \\
        &      &        & 86.32(6.96)     & 527(10) \\
\net  & 15.56 & 41.0    & 59.48(2.88)     & 176(4) \\
        &      &        & 188.46(9.11)    & 529(11) \\
\nath  & 7.32  & 47.3   & 0.94(0.19)      & 200(1) \\
        &      &        & 2.44(0.50)      & 479(86) \\
\hline
\end{tabular}}
\tablecomments{Lines with more than one component have multiple rows. }
\end{table}

\subsection{Coronal Gas}

Coronal lines are high-ionization emission lines that were first detected in the Sun's corona. The term is commonly used to define lines with IPs greater than 100\,eV, and are often found in AGN (see \citealp{Rodriguez2025} for a review). Indeed, these lines have been shown to reliably trace AGN activity \citep{RodriguezArdila2002, Kynoch2022}. 

In this paper, we define coronal lines based on the log(U) value at which the corresponding ionic abundance peaks \citep{feuillet:etal:footprint:2026}. More specifically, the coronal lines peak at a log(U) $\approx$ -1.25 (see Figure \ref{fig: abundances}) and include lines all with IPs greater than that of the He \textsc{ii} Lyman limit (54.4\,eV), ranging from 54.9 to 126.2\,eV. Their fluxes and FWHM values are given in Table \ref{tab:high_lines}.

\begin{center}
\begin{table}[h!]
\centering
\caption{Summary of the coronal and footprint lines: best-fit fluxes and FWHM.}
\label{tab:high_lines}
\resizebox{\columnwidth}{!}{%
\begin{tabular}{lcccc}
\hline
Line & $\lambda_{rest}$ & IP  & Flux & FWHM \\
 & (\mic) & (eV) & (10$^{-14}$ erg\,cm$^{-2}$\,s$^{-1}$) & (km\,s$^{-1}$) \\
\hline\hline
\multicolumn{5}{l}{Coronal Lines} \\
\hline
\of    & 25.89 & 54.9   & 21.78(1.75)     & 183($<$1) \\
        &      &        & 178.86(14.37)   & 512(15) \\
\arv & 7.90  & 75.0     & 2.83(0.64)      & 210(100) \\
        &      &        & 2.63(0.29)      & 587(280) \\
\arv  & 13.10 & 75.0    & 2.19(0.40)      & 210(100) \\
        &      &        & 4.58(0.84)      & 587(280) \\
\nev &	14.32 &	97.1   & 32.80(2.28)      & 165(2) \\
     &        &        &	67.00(4.67)      & 553(9) \\
\nev &	24.32 & 97.1   & 14.80(2.35)      & 165(2) \\
     &        &        &	45.50(7.21)	     & 553(9) \\
\fesev  & 7.82  & 99.1  & 3.29(0.66)      & 248(11)      \\
        &      &        &  3.23(0.64)     & 530(169)      \\
\fesev  & 9.53  & 99.1  & 5.14(1.43)      & 248(11)      \\
        &      &        &  7.47(2.08)     & 530(169)      \\
\mgv    & 5.61  & 109.2 & 7.72(0.60)      & 194(8)\\
        &      &        & 12.60(0.98)     & 537(59) \\
\nesix   & 7.65  & 126.2 & 63.90(2.31)    & 212(3)      \\
        &      &         & 53.90(1.94)    & 524(25)      \\
\hline
\multicolumn{5}{l}{Footprint Lines} \\
\hline
\feeig & 5.45  & 124.0 & 3.70(0.40)       & 207(3)      \\
        &      &       & 9.31(0.99)       & 441(39)      \\
\mgsev  & 5.50  & 186.5 & 4.97(0.65)      & 182(9)      \\
        &      &        &  8.13(1.06)     & 382($<$1)      \\
\sisev  & 2.48  & 205.3 & 2.09(0.72)      & 124(10)      \\
        &      &        & 4.51(1.55)      & 538(177)      \\
\mgeig & 3.03 & 225.2 & 1.69(0.75)        & 124(23)      \\
        &      &        &  2.65(1.18)     & 161(76)      \\
\sinine & 2.58 & 303.2 & 0.77(0.15)       & 200(1)      \\
        &      &        & 1.16(0.23)      & 452(87)  \\   
\hline
\end{tabular}}
\end{table}
\end{center}

\subsection{Footprint Gas}

As previously introduced, "footprint lines" are high-ionization emission lines that can be used to trace X-ray gas, as they have been shown to be produced within similar gas conditions. This method was first introduced by \cite{trindade_falcao_tracking_2022} and used NGC~4151 as the target for demonstration. However, the authors were mainly focusing on one optical footprint line ([Fe~\textsc{x}]\,6375\,\angstromns) with HST/STIS data.

The emission lines proposed as footprint line candidates by \cite{trindade_falcao_tracking_2022} that are also present in our data are \mgsev~5.50\micns, \sisev~2.48\micns, \mgeig~3.03\micns, and \sinine~3.55\micns. Based on Figure~\ref{fig: abundances}, we identify footprint lines as those whose corresponding ionic abundances show significant overlap with the X-ray abundances. This selection includes the lines identified by \cite{trindade_falcao_tracking_2022}, as well as \feeig~5.45\mic, which was also classified as a footprint line in NGC~7469. The observed X-ray emission lines and their corresponding IR footprint lines are as follows:

\begin{itemize}[leftmargin=*]
    \item \osev~$\leftrightarrow$~\feeig~$\lambda$5.45\micns,~\mgsev~$\lambda$5.50\micns;
    \item \csixB $\leftrightarrow$ \mgeig $\lambda$3.03\micns,
    \item \neninens, \nitsev $\leftrightarrow$ \sinine $\lambda$2.58\micns.
\end{itemize}

Additionally, as found in NGC 7469, the X-ray line \nitsix peaks at log(U) values similar to those of the coronal lines. Thus, although \nesix and \sisev technically qualify as footprint lines because they trace the same gas as an X-ray line, \nitsix does not probe the same extremely high-ionization gas traced by the other X-ray lines. Instead, \nitsix traces coronal gas already characterized at optical wavelengths, which is not the focus of this work. Our interest lies in the spatially resolved X-ray-emitting gas that can only be accessed through footprint lines in combination with JWST observations.

The fact that the \sfo and the \feeig span a similar velocity range in the profiles shown in Figure \ref{fig:profile_comparison} suggests that the two components may be interacting, potentially through entrainment or confinement. A more detailed analysis of the kinematics and dynamics of the different line categories will be presented in Paper II, where this phenomenon will be investigated in a spatially resolved manner to better understand the underlying physical processes.


\section{Global Photoionization Modeling} \label{sec:global_modeling}

\begin{figure*}[]
\centering
\includegraphics[width=\textwidth]{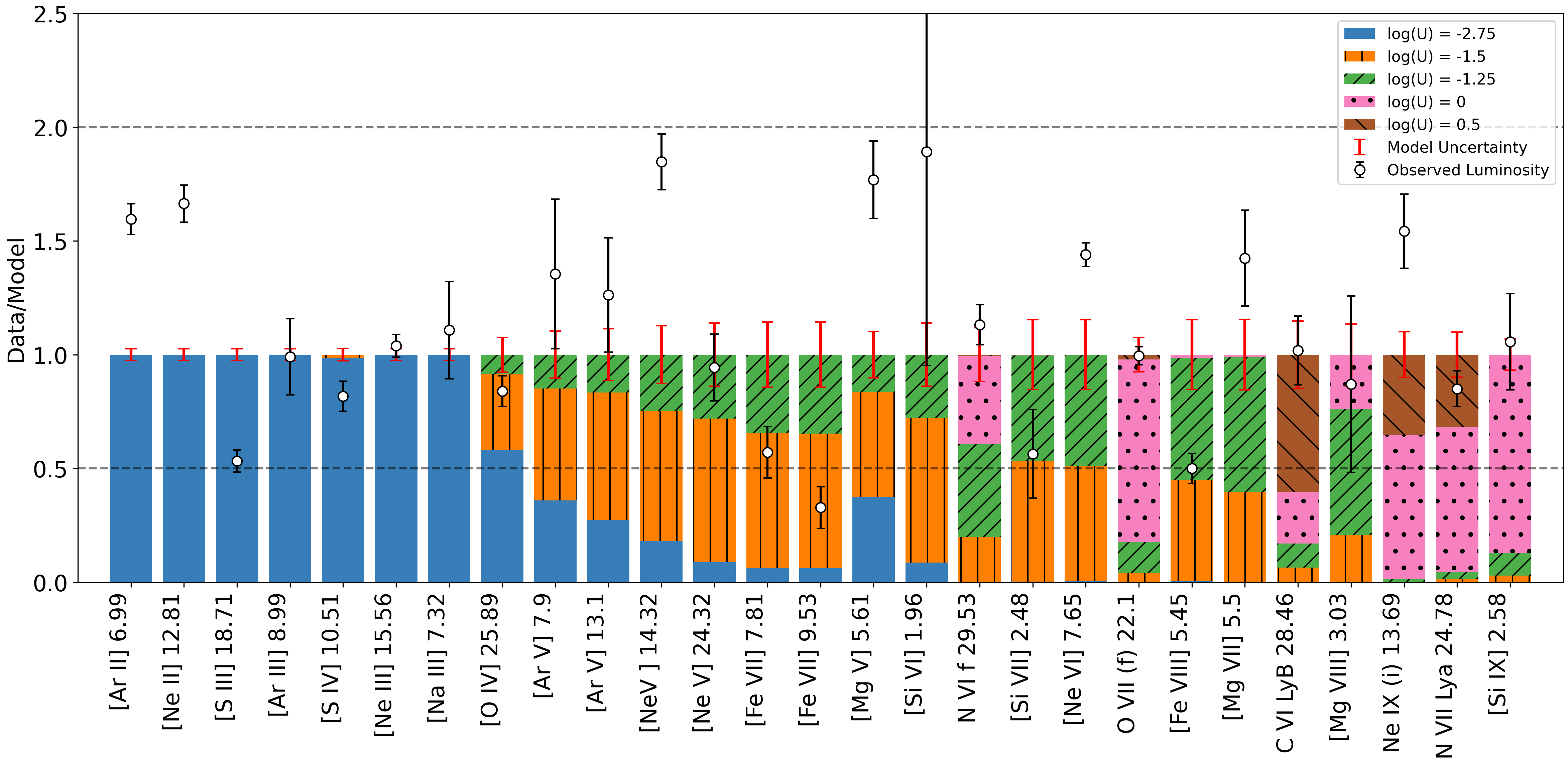}
    \caption{Ratios of observed to predicted luminosities of each relevant emission line in the NGC~4151 global spectrum, in order of increasing IP. The bars represent the predicted luminosities, broken down by the contribution of each \texttt{Cloudy} component, with colors as indicated in the legend. The white circles are the ratios of observed to predicted luminosities and the dotted lines indicate a factor of 2 difference between the observed and modeled values.  
    \label{fig: global_models}}
\end{figure*}

In order to perfectly fit all of the emission lines included in the previous two tables, considering the wide range in their IPs, and ion abundance peaks, one would need a similarly wide range of combined \Cloudy models. While some have argued that this is necessary to successfully model the NLR of AGN (e.g. the Locally Optimally emitting Cloud model by \citealp{Ferguson1997}), many studies have shown that this can be done with satisfactorily accuracy using a smaller number of components \citep{armentrout_photoionization_2007, kraemer_mass_2020}.

To construct the photoionization models, we first adopted the ionization potentials listed in \cite{armentrout_photoionization_2007} with log(U)\,=\,1.3, 0, and -0.5 and column densities limited by the size of our region. These initial models were then used to fit both the footprint emission lines and the X-ray emission lines simultaneously. We use an average deprojected distance from the nucleus of 72\,pc (or 4.5 spaxels, \citealp{meena_investigating_2023}) and Equation \ref{eq: U} to calculate the hydrogen density for each model (see \citealp{feuillet:etal:footprint:2026} for more details on the fitting process). During this fitting, the component with log(U)\,=\,1.3 returned a covering factor of 1 which is unphysical. This occurred as it was the only component capable of producing \csixB emission but with a small modeled flux, it required the model to compensate by increasing the emitting area. As discussed in \cite{armentrout_photoionization_2007}, this highly ionized component is primarily intended to reproduce higher-ionization species such as O \textsc{viii} Ly$\alpha$ and Ne \textsc{x} Ly$\alpha$, which are outside the scope of the present analysis, but could be traced by optical footprint lines such as [Fe~\textsc{xiv}]~5303\,\angstromns.

To address this issue, we replaced the log(U)\,=\,1.3 component with one at log(U)\,=\,0.5. This modification significantly improved the fits to the observed lines while also eliminating the unrealistically large covering factor. The resulting high-ionization three-component model consisted of components at log(U)\,=\,-0.5, 0, and 0.5.

We additionally ran \Cloudy models with log(U) values ranging from -3 to 1 in steps of 0.25, and column densities from log(N$_H$/cm$^{-2}$)\,=\,20.5 to 22 to fit the rest of the IR lines. Note that all of the emission lines in Tables \ref{tab:low_lines} and \ref{tab:high_lines} are used in the fitting routine except the H recombination lines and the \fetw lines, as they may have different origins than the other emission lines. These non-footprint lines are best reproduced using components at log(U)\,= -2.75, -1.5, and -1.25. These components are consistent with previously used photoionization models of NGC~4151 in \cite{Kraemer2000_PA} (log(U)\,=\,-1.1 and -2.7) to fit optical data.


Combining both sets and performing a final fit, we find that both the IR and X-ray lines are best reproduced using the following five components: log(U)\,=\,\mbox{-2.75}, -1.5, -1.25, 0, 0.5 (\texttt{nebular}, \texttt{coronal1}, \texttt{coronal2}, \texttt{footprint1}, \texttt{footprint2}) with respective hydrogen densities of log(n$_H$/cm$^3$)\,=\,3.96, 2.71, 2.46, 1.21, 0.71 and column densities of log(N$_H$/cm$^2$)\,=\,20.5, 21.8, 22.2, 21.8, and 21.3.  The results for our 5 component model are shown in Figure \ref{fig: global_models}, where the closer the normalized observed luminosity is to 1, the better the model matches observations. 

All of the emission lines models, including those of the X-ray lines, agree with the measured values within a factor of 2 within the corresponding error bars. The only exception is the \fesev 9.53\mic line, which is overpredicted by a factor of 3. This discrepancy likely stems from the line fitting as this emission line is in a region of strong fringing, likely affecting our measurements. Note that the fit for the other \fesev line is acceptable.

Based on our models, we can calculate the mass of the global mass for each model component. A detailed explanation of the procedure is given in \cite{feuillet:etal:footprint:2026}, with a short outline given here. The emission lines above were fitted with two constraints: the chosen model must not exceed the column of gas available, or the maximum emitting area at that distance. Once the best fit component set is chosen to fit the emission lines, the resulting values are then used to calculate the masses as such:
\begin{equation}\label{eq: Mass}
    M = n_{H} A \delta \mu m_p
\end{equation}
where $\delta$ is the thickness and A is the area of the emitting gas, $n_H$ is the Hydrogen density, $\mu$ is 1.3, and $m_p$ is the mass of a proton.

From Equation \ref{eq: Mass} and the information presented in Table \ref{tab: Masses}, we obtain an X-ray gas mass of (1.4\,$\pm$\,0.2)\,$\times$\,10$^{5}$\,M$\odot$⁠, in comparison to (7.8\,$\pm$\,2.1)\,$\times$\,10$^{5}$\,M$\odot$⁠ from the previous footprint analysis \citep{trindade_falcao_tracking_2022}, and  (5.4\,$\pm$\,1.1)\,$\times$\,10$^{5}$\,M$\odot$ measured from a Chandra X-ray analysis \citep{kraemer_mass_2020}. 


Our estimate is smaller than the previously measured values as it is based on an assumed average distance, whereas the previous values were derived from spatially resolved data. This is a significant difference as global parameters can vary by up to three orders of magnitude within a single target due to their strong dependence on filling factor and gas density, and because they do not account for the spatial distribution of the outflowing material \citep{Revalski2018, Revalski2022}. Calculating mass and energy outflow rates using spatially resolved data, as done in the study of NGC 7469 \citep{feuillet:etal:footprint:2026}, offers the most reliable way to obtain accurate and well-constrained values. Therefore, we expect to place a stronger constraint on the X-ray gas mass in our forthcoming paper.

\section{Discussion} \label{sec: discussion}

NGC~4151 is an archetypal Seyfert~1.5 AGN with a well-defined biconical outflow, making it an ideal target for studying AGN feedback. The broad wavelength coverage of JWST enables a comprehensive, multiphase view of the outflow and surrounding medium, allowing direct comparisons between the warm molecular gas, the nebular gas, traditionally traced at optical wavelengths with \otns, and the X-ray-emitting phase.

\subsection{A Multi-Phase Mass Budget}

When assessing the potential impact of the AGN on its host galaxy, the two primary quantities to consider are the mass of the outflowing gas and its kinematics, as both enter directly into the calculation of the mass outflow rate. In this paper, we estimated the masses of both the molecular and ionized gas components, and the resulting values are summarized in Table~\ref{tab: Masses}. As shown by our calculations, the largest gas reservoir resides in the cold molecular phase, followed by the warm molecular gas. Among the ionized gas phases, however, the X-ray-emitting gas traced by the footprint lines is most massive.

As shown in Figure~\ref{fig:profile_comparison}, emission lines spanning a wide range of ionization potentials all display an outflow component centered at approximately -250\,km\,s$^{-1}$, indicating a common kinematic structure across the different gas phases. On the other hand, the molecular gas does not exhibit a significant kinematic outflow component in our observations. We additionally do not detect molecular gas associated with the ionization bicone, and therefore find no compelling evidence for a molecular outflow in our data.

\begin{center}
\begin{table}[]
\centering
\caption{Calculated gas masses for the different phases of the outflow.}
\begin{tabular}{ccccc}
\hline
Gas & log U & Area & $\delta$ & Mass\\
 &  & (10$^{41}\,$cm$^2$) & (cm) & (10$^{4}$\,M$_{\odot}$)\\
 \hline
 \hline
Ionized & & & & \\
\hline
Nebular & -2.75 & 2.37$\pm$0.06 & 3.40E+16 & 0.0035    \\
\multirow{2}{*}{Coronal} & -1.5  & 1.36$\pm$0.19 & 1.35E+19 & 0.55    \\
& -1.25 & 1.73$\pm$0.29 & 5.34E+19 & 2.5    \\
\multirow{2}{*}{Footprint} & 0 & 0.88$\pm$0.05 & 3.91E+20 & 4.5    \\
 & 0.5 & 3.17$\pm$0.57 & 3.91E+20 & 9.6  \\
\hline
Molecular & & & & \\
\hline
T $>$ 200\,K  & - & - & - & 47\\
T $>$ 50\,K & - & - & - & 1800\\
\hline
\end{tabular}
\label{tab: Masses}
\end{table}

\end{center}

Since the X-ray-emitting gas contains the largest mass among the outflowing ionized phases while exhibiting a comparable outflow velocity, it is expected to dominate the mass outflow rate and, consequently, have the greatest impact on the host galaxy. This was suggested in \cite{kraemer_mass_2020}, but the lack of opportunity for direct comparison prevented any further conclusions. Note, however, that due to the previously mentioned issues with obtaining global mass and mass outflow rate values, a more accurate breakdown of the impact of each phase on the host galaxy will be possible with the spatially resolved investigation in Paper II.

\subsection{Evolutionary Implications}

In \cite{Trindade2026}, the authors propose an evolutionary sequence in which quasars enter a post-blowout phase characterized by extended optical and X-ray winds. NGC~4151 exhibits well-defined biconical structures in both the nebular and footprint emission lines, consistent with this scenario. This was also shown in Figure \ref{fig: chandra_soft} where the \ot bicone aligns perfectly with the soft X-ray emission. Another predicted signature of the post-blowout phase is kinematic coupling between the X-ray wind and the optical narrow-line region, which matches up with our observations as, as aforementioned, we found a common kinematic structure across the different gas phases when comparing the global line profiles.

Together, the morphology and kinematics of the nebular and X-ray gas suggest that NGC~4151 is in a post-blowout stage of AGN evolution. This interpretation is further supported by the apparent absence of nuclear SF, implying that the ISM gas required for SF has been evacuated or disrupted by the outflow.

The combination of suppressed star formation and the strong coupling between the outflow and the host galaxy is consistent with generalized negative feedback, where AGN-driven winds inhibit SF. Confirming both the absence of SF and the efficiency of this feedback, however, requires a spatially resolved analysis. In the second paper of this series, we will measure the feedback efficiency, defined as the ratio of the maximum energy outflow rate to the bolometric luminosity \citep{Hopkins2010}, and quantify the contribution of each gas phase to the total feedback. We will also place individual spaxels on the \cite{Feuillet2025} infrared diagnostic diagrams to search for localized star-forming regions that may be diluted in integrated measurements.

The spatially resolved analysis will also enable a direct comparison with the results of \cite{feuillet:etal:footprint:2026} for NGC~7469. Despite their similar bolometric luminosities, the two galaxies exhibit markedly different outflow morphologies. \cite{Fischer2013} classified NGC~7469 as having a compact morphology and NGC~4151 as exhibiting an outflow morphology. In \cite{feuillet:etal:footprint:2026} we attributed the compact \ot and X-ray morphology of NGC~7469, together with its prominent nuclear star-forming ring, to the orientation of the ionization bicone. In the context of the evolutionary sequence proposed by \cite{Trindade2026}, an alternative interpretation is that the AGN in NGC~7469 is in a pre-blowout stage. The spatially resolved analysis of NGC~4151, to be presented in Paper II, will enable a more comprehensive comparison between these two systems.

\section{Conclusions}

In this paper, we use new JWST data to study the multiphase outflow in NGC~4151. We present a summary of our results below.

\begin{enumerate}
\item Using \Cloudy photoionization modeling, we determined where the ions corresponding to the main emission lines peak in terms of ionization parameter, allowing us to categorize them and identify which lines reliably trace the X-ray outflows in NGC~4151.

\item We used the range of H$_2$ lines available in the MIRI data to calculate a total H$_2$ mass of $\sim$4.7\,$\times$\,10$^5$\,M${\odot}$ between 200 and 2000\,K, and $\sim$1.8\,$\times$\,10$^7$\,M${\odot}$ between 50 and 2000\,K within the MIRI FOV. Although the molecular gas and \fetw\ emission can sometimes share a common origin, we find that the H$_2$ gas is concentrated in the cross-cone region, whereas the \fetw\ emission is associated with the bicone along the radio jet.

\item Across a broad range in IP, the global line profiles consistently exhibit an outflow component at \mbox{$\sim$-250\kmsns}, indicating a common kinematic structure.

\item Additional \Cloudy\ modeling allowed us to characterize the heterogeneous gas in NGC~4151 with a five-component model spanning more than three orders of magnitude in log(U), and thus density.

\item By calculating the total mass of the different gas components, we find that the X-ray-emitting gas is the most massive ionized gas phase. This suggests that it is likely the dominant contributor to AGN feedback in the host galaxy.

\end{enumerate}

The second paper in this series will leverage the full spatial capabilities of JWST to address the spatially resolved properties of the gas in NGC~4151.

\begin{acknowledgments}

This work is based on observations made with the NASA/ESA/CSA James Webb Space Telescope. The data were obtained from the Mikulski Archive for Space Telescopes at the Space Telescope Science Institute, which is operated by the Association of Universities for Research in Astronomy, Inc., under NASA contract NAS 5-03127 for JWST. The specific observations analyzed can be accessed via MAST DOI: DOI will be added when abstract is finalized. RM acknowledges financial support from the INAF Scientific Directorate.

\end{acknowledgments}





%
\facilities{JWST - James Webb Space Telescope (MIRI), JWST - James Webb Space Telescope (NIRSpec), XMM - Newton X-Ray Multimirror Mission satellite (RGS), CXO - Chandra X-Ray Observatory (HETG).}



\appendix

\section{PSF Subtraction} \label{app: PSF_sub}

As AGN are very bright, point-like sources, the central emission is often subject to PSF issues, particularly in Type 1 AGN like NGC~4151 and NGC~7469, where the emission that would be contained in the center-most spaxel is spread out over the FOV in a specific pattern based on the geometry of the instrument. To mitigate this issue, PSF simulations or template star observations can be used to remove the excess emission.

The first method involves using \texttt{STPSF}, a Python software that produces PSF simulations for JWST to create our PSF cubes \citep{Perrin2012, Perrin2014_PSF, Perrin2025-zenodo}. The code allows for a customizable spaxel size, which we set to 0\farcs196 to match our data. As outlined in the data analysis section, we use the cubes processed with the \texttt{ifualign} orientation in order to do the subtraction without needing to rotate the PSF, which could introduce numerical interpolation noise \citep{Perrin2025-zenodo}. 

The second method makes use of an observed template star, another point source, to get an empirical shape of the PSF and subtract it from the data. For NGC~4151, the template stars observed alongside the NIRSpec observations cannot be used for our purposes as they did not use the same instrument configuration as the NGC~4151 ones. On the other hand, the template star observed with the MIRI instrument was specifically obtained for PSF subtraction purposes. We downloaded and processed the template star observations in the JWST pipeline in the same manner as the data (see Section \ref{sec:IR_data}). 

\begin{figure*}[h!]
\centering
\includegraphics[width=\textwidth]{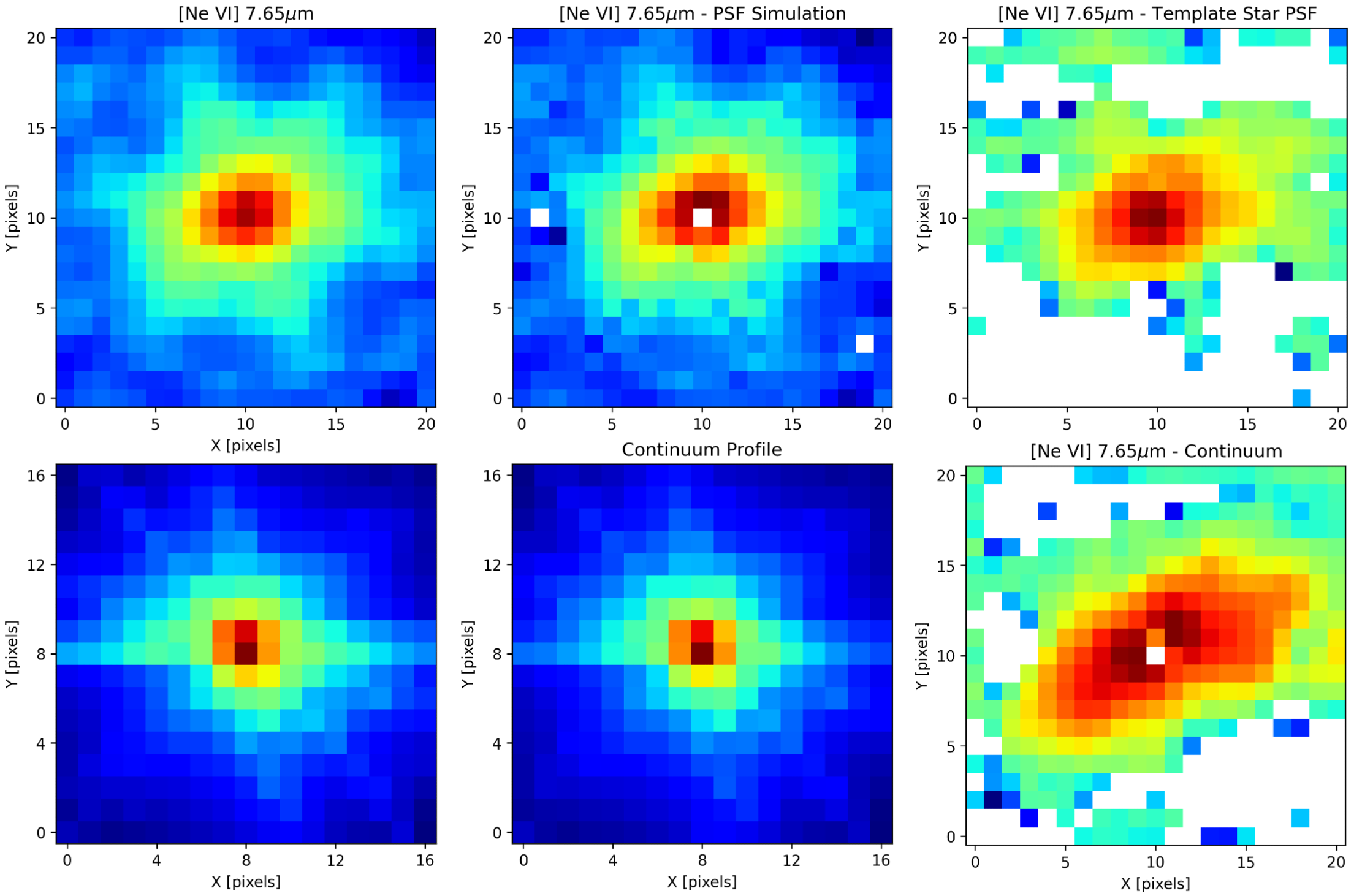}
    \caption{PSF and continuum subtraction examples. The top left panel shows the \nesix distribution directly from the pipeline output showing the PSF structure. The top center and right panels demonstrates the PSF subtraction using the \texttt{STPSF} simulation and template star respectively, with the residual PSF. The bottom left and center panels show the distribution of the broad Pa$\alpha$ line and nearby continuum emission, establishing their similarity. The bottom right panel shows the continuum subtracted \nesix image, with the PSF wings absent from the result. 
    \label{fig:PSF}}
\end{figure*}

We attempted both PSF subtraction methods for our data, and in each case first aligned the peak of the PSF cubes to the peak in the NGC~4151 emission for each individual frame. We then scaled the PSF cube to remove different percentages of the central spaxel. The subtraction was unsuccessful in both cases: the PSF simulations were not perfectly representative of the observed PSF, leading to residual PSF, and the template star was not bright enough to produce a strong PSF across all channels (see Figure \ref{fig:PSF}). 

\subsection{PSF Subtraction vs. Continuum Subtraction}

If the continuum emission originates solely from the AGN, rather than from any extended stellar or nebular components, the PSF can be effectively removed through continuum subtraction. In this scenario, the continuum traces only the unresolved central source, allowing the subtraction process to isolate the extended emission. Forbidden emission lines are expected to arise from gas distributed on large spatial scales, and as this emission is intrinsically extended, it is not significantly affected by the PSF and remains after continuum subtraction. 

Additional evidence that the continuum is dominated by the central engine comes from the spatial distribution of the broad component of the Pa$\alpha$ emission line. Because the broad Pa$\alpha$ emission originates in the unresolved BLR close to the AGN, it should exhibit the same spatial profile as the continuum if both arise from the same central source (see Figure \ref{fig:PSF}). The observed agreement between the broad Pa$\alpha$ and continuum distributions therefore demonstrates that the continuum emission is entirely consistent with originating from the central engine, validating the use of continuum subtraction to remove the PSF contribution.

\section{Footprint Line Morphology and Kinematics} \label{app: kinematics}

As shown Section \ref{sec: modeling}, if the X-ray lines and their corresponding footprint lines are produced in the same gas, both the morphology of their emission and their global kinematics should be similar. This was demonstrated in \cite{feuillet:etal:footprint:2026} for NGC 7469, however, the X-ray outflows for that target are compact and no conclusive results were able to be drawn from the comparison beyond a general agreement. 

\begin{figure*}[]
\centering
\includegraphics[width=\textwidth]{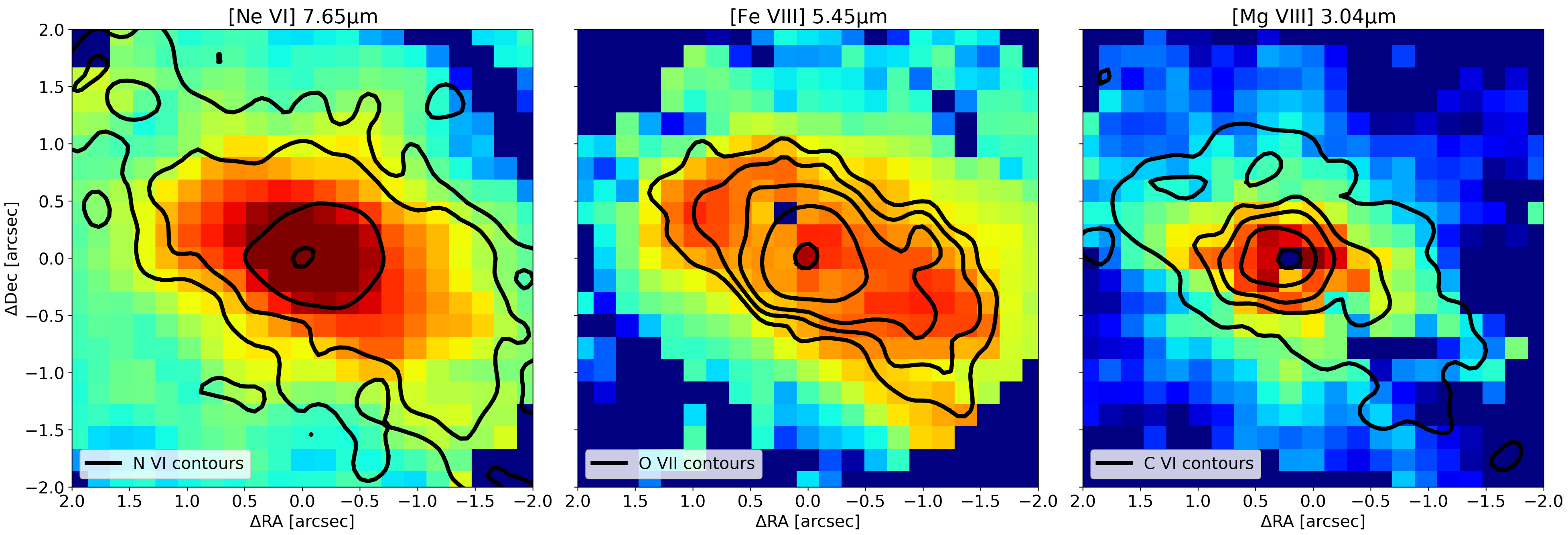}
    \caption{Each panels shows IR line maps for \nesixns, \feeigns, and \mgeig and the contours from their corresponding X-ray lines from the Chandra data described in Section \ref{sec: Chandra}. The Chandra images are PSF-subtracted as outlined in the text and smoothed with a Gaussian kernel ($\sigma$ = 1.5, r = 3), and the contours level are [1, 1.4, 2, 4, 12] for \osevns, [0.2, 0.5, 1.5, 6] for \nitsixns, and [0.3, 0.8, 1.2, 2, 2.7] for \csix.
    \label{fig: X-ray contours}}
\end{figure*}

In Figure \ref{fig: X-ray contours}, we show the IR maps obtained from \texttt{QFitsView} \citep{Ott2012} for the emission lines of interest, where we use the sky align data cubes for easy comparison to the X-ray data. These data are not PSF subtracted, but are continuum subtracted (see Appendix \ref{app: PSF_sub} for a discussion). We find that the X-ray and corresponding footprint line morphologies agree well across all three examples. 

The north-east side of the bicone in this target is pointed into the plane of the host galaxy, likely attenuating the X-ray emission due to intervening gas. This presents an issue when attempting to compare the kinematic properties of the global X-ray and footprint lines. Indeed, when studying the X-ray line properties, the sampled gas will be primarily originating from the south-west region of the bicone (Figure \ref{fig: chandra_soft}). On the other hand, a global value for the IR footprint lines will include both the blueshifted and redshifted gas, as evidenced by the symmetrical profiles of the lines in Section \ref{sec: outflow}. To mitigate this issue, we also fit the emission lines in the SW cone region as defined in Figure \ref{fig: regions} to facilitate the comparison to the global X-ray fluxes.

\begin{figure}[h!]
\centering
\includegraphics[width=0.45\textwidth]{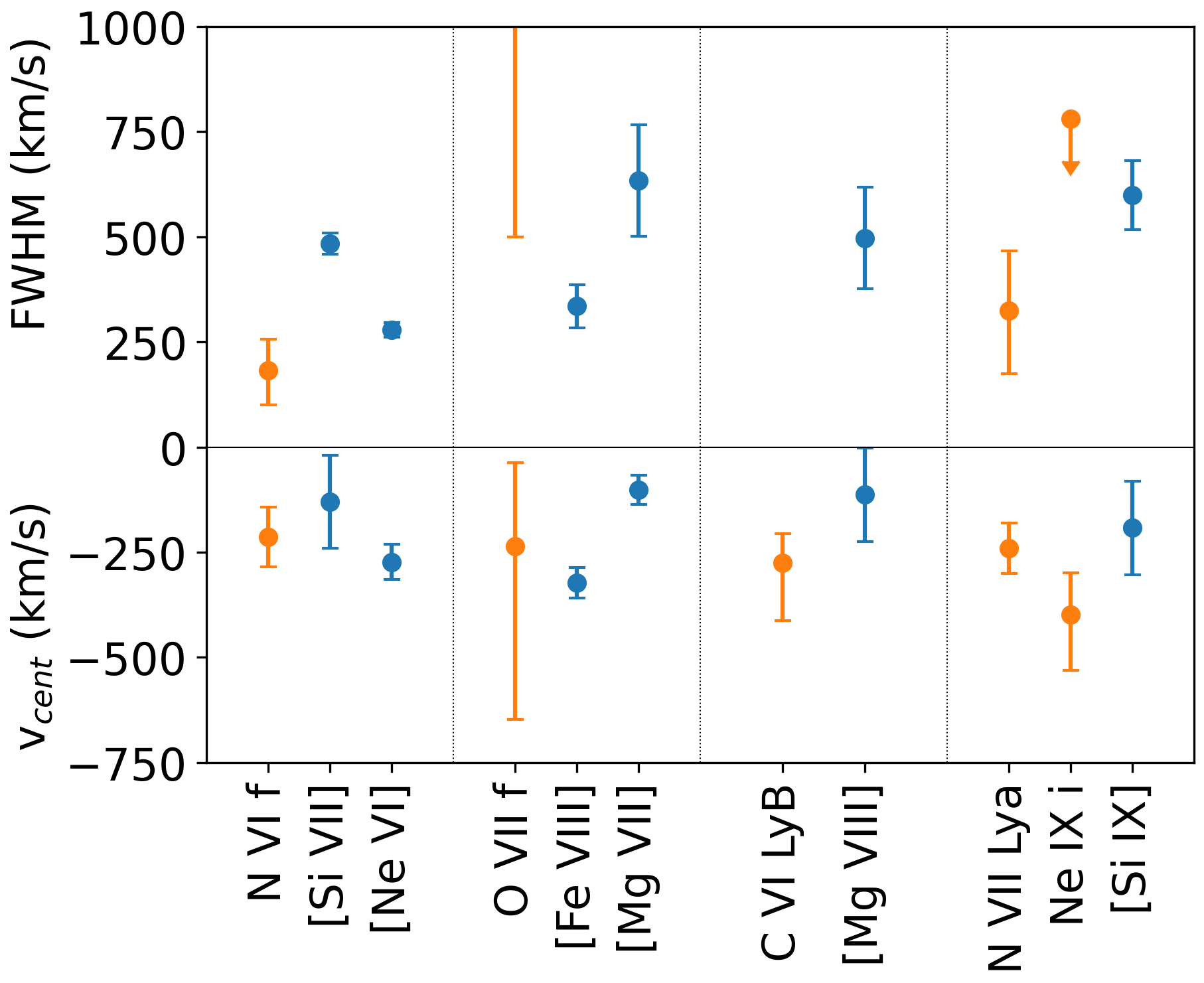}
    \caption{This plot shows the FWHM and v$_{cent}$ values for both the X-ray lines as compiled in Table \ref{tab: x-ray lines}, and the blueshifted component of the corresponding IR footprint lines. As discussed in the text, the IR lines are fitted from the SW cone region rather than the full region. 
    \label{fig:IR_xray_comparison}}
\end{figure}

The fit results for the blueshifted component of the IR lines and the comparison to their corresponding X-ray lines are shown in Figure \ref{fig:IR_xray_comparison}. The lines are ordered by increasing log(U) peak as determined in Figure \ref{fig: abundances}. There is a general trend of increasing FWHM when moving towards the gas peaking at higher log(U) values, which is not present in the v$_{cent}$ values.

We found that the kinematic properties of the footprint lines are consistent with those of the X-ray lines. This result is important for future work, as the footprint lines provide substantially more kinematic information than the X-ray lines due to the spatial and spectral resolution of JWST. We confirmed that the physical conditions are globally consistent between the two sets of lines, which will enable more accurate determinations of outflow parameters in future studies using spatially resolved data.

\bibliography{sample701}{}
\bibliographystyle{aasjournalv7}



\end{document}